\documentclass[12pt]{article}

\usepackage{graphics}
 
\begin{document}
 
\title{Effects of Flow on Measurements of \\ Interactions in Colloidal Suspensions}
 
\author{Yuri O.\ Popov\thanks{E-mail: {\tt yopopov@midway.uchicago.edu}}}
\date{{\em Department of Physics, University of Chicago,} \\
{\em 5640 S.\ Ellis Ave., Chicago, IL 60637, USA}}
\maketitle 

\begin{abstract}

A hydrodynamic mechanism of interactions of colloidal particles is considered.  The mechanism is based on the assumption of tiny background flows in the experimental cells during measurements by Grier {\em at al.}~\cite{crocker1, crocker2, grier1, larsen1, crocker3, grier2}.  Both trivial (shear flow) and non-trivial (force propagation through viscous fluid) effects are taken into account for two colloidal particles near a wall bounding the solvent.  Expressions for the radial (attractive or repulsive) forces and the polar torques are obtained.  Quantitative estimates of the flow needed to produce the observed strength of attractive force are given, other necessary conditions are also considered.  The conclusion is made:  the mechanism suggested most likely is {\em not\/} responsible for the attractive interactions observed in the experiments of Grier {\em at al.}; however, it may be applicable in other experimental realizations and should be kept in mind while conducting colloidal measurements of high sensitivity.  Several distinctive features of the interactions due to this mechanism are identified.

\end{abstract}

\begin{center} {\bf PACS}: 82.70.Dd --- Colloids; 47.15.Gf --- Low-Reynolds-number (creeping) flows \end{center}

\section{Introduction}

The motivation for this work comes from the experiments conducted by David Grier {\em at al.}~\cite{crocker1, crocker2, grier1, larsen1, crocker3, grier2} as well as by other groups~\cite{kepler, carbajal}, which show unexplained attraction between colloidal particles in suspension.  As most other polymer aggregates do, the particles used in the experiments dissociate in the solvent with charges of one sign remaining on the surface of the particles and charges of the other sign going to the solvent.  According to the prevailing model for electrostatic interactions of such particles (DLVO~\cite{derjaguin, verwey, russel}), these particles of same-sign charge should repel via a screened Coulomb potential.  Several other theories for electrostatic interaction mechanisms have been suggested to explain the attraction observed~\cite{sogami, tata, allahyarov1, allahyarov2}, but none of them stood the experimental tests.   Non-electrostatic models have not succeeded in providing concrete reasons for the observed behavior either.  However, the first hydrodynamic attempt to account for the attraction in the recent work of Squires and Brenner~\cite{squires} appears quite successful.  

Here we explore another hydrodynamic effect that complements the effect described by Squires and Brenner.  The nature of attractive force is essentially the same (a non-equilibrium hydrodynamic effect based on the method of image forces by Blake~\cite{blake}), but the mechanism explored here does not require any charges on the surface of the walls and colloidal particles, while the mechanism of Squires and Brenner does.  While we do not intend to dismiss their mechanism and we agree that it quantitatively captures essential features of the experimentally observed attraction, we would like to offer a complimentary effect that might be responsible for attraction in the cases when the explanation of Squires and Brenner does not apply.  Examples of such systems are suspensions and meta-stable crystallites, where particles are essentially fixed in space (i.e.\ fluctuate around their equilibrium position).  This limits the mobility of particles, which decreases applicability of the mechanism of Squires and Brenner, but increases applicability of ours (as being ``fixed in space'' is an essential condition for our mechanism).  While quantitative description of suspensions and meta-stable crystallites requires further work, we feel that the mechanism described below is potentially important in accounting for colloidal interactions and complements the picture drawn by Squires and Brenner.  It also demonstrates a subtlety produced by undetectable hydrodynamic flows that is relevant for colloidal measurements.  However, this mechanism appears too weak to account for the experiments that motivated its development as we currently understand them.

In this note we first consider physical ideas and assumptions behind this mechanism, and then we theoretically investigate two hydrodynamic effects resulting from these assumptions.  In the discussion section that follows, we estimate orders of magnitude of these effects and conclude that the mechanism explored is probably too weak in the particular conditions of the experiments of Grier {\em at al.}  Later we discuss possible other implications and experimental tests.

\section{Physical assumptions and geometry}
 
The experiments of Crocker and Grier were conducted on the samples of colloidal suspension confined between a microscope slide and a cover slip.  Each experimental run consisted of a series of recordings of Brownian motion of two colloidal particles (sufficiently remote from the other particles in the suspension) and subsequent analysis of these recordings, which allowed one to infer the pair interaction potential~\cite{crocker1, crocker2}.  Before each recording the particles were caught in the focal plane of the digital video microscopy setup by optical tweezers, which are essentially a potential well of the electromagnetic origin~\cite{ashkin, grier3}, then they were released, and their Brownian motion was recorded in several $1/30$-of-a-second periods.  Subsequently the particles were trapped by the tweezers again (in order to be returned to the recording area of the sample) and a new cycle began.

Here we consider the possible effects of background flow in the apparatus.  Our motivation originated from the thorough flushes of the solvent carried out with the purpose of removing impurities from the solution before each experimental run.  However, the flow induced by these flushes decays exponentially with characteristic time of the order of $d^2/8\nu$, where $d$ is the distance between the slide and the cover slip and $\nu$ is the kinematic viscosity of the liquid.  Hence the flushes cannot be a source of sustained flow:  the typical value of this time in the experimental conditions is of the order of $10^{-4}$--$10^{-3}$~s.  At the same time there is another effect capable of inducing some flows that are both small enough to be experimentally undetectable and large enough to produce the desirable attractive effect during the entire experiment (the required magnitude of flow velocity will be shown to be of the order of $0.1\mbox{ }\mu\mbox{m}/\mbox{s}$).  This effect is related to a slight misbalance in pressure in a vacuum system connected to the experimental cell.  Each cell was connected to the colloid suspension supply via two openings in the microscope slide.  Colloid refill, ion removal and all other auxiliary operations were conducted via this system.  When measurements of the attractive potential were performed, some negative pressure was applied inside the cell to make the cover slip bow inwards, lowing wall separation $d$ from about 50~$\mu$m to about 5~$\mu$m.  While the pressure inside the vacuum system was of the order of 15~Torr or 2000~Pa, we estimate\footnote{The only non-trivial component of the Navier-Stokes equation reads $\frac{\partial p}{\partial x} = \eta \frac{\partial^2 u}{\partial z^2}$, with $\hat x$ being along the flow and $\hat z$ being normal to the walls ($\eta$ is dynamic viscosity).  Since pressure $p$ does not depend on $z$, the standard solution is the parabolic profile for velocity $u = \gamma z (d-z)/d$, with $d$ being the distance between the parallel walls.  This yields $|\Delta p| = 2 \eta \gamma |\Delta x|/d$ for pressure difference between the openings separated by $\Delta x$.  Typical values $\Delta x = 2\mbox{ cm}$, $\eta = 1\mbox{ mPa}\cdot\mbox{s}$, and $d = 5\mbox{ }\mu\mbox{m}$ give a result $|\Delta p| = 0.8\mbox{ Pa}$ for $\gamma = 0.1\mbox{ s}^{-1}$.} that the difference in pressure between the two parts of the vacuum system had to be only 0.8~Pa in order to produce the required magnitude of flow.  Thus, 0.04\% pressure difference, which is definitely beyond the experimental control and can exist unnoticeably in the conditions of the experiment~\cite{grier4}, could cause the background flow inside the cell.  In general, since the required flows are very small, other subtle effects can also contribute to their existence.  All the subsequent treatment is based on the assumption of existence of small constant flows in the experimental cell.  One of the purposes of this paper is to show that these undetectable flows can lead to quite detectable effects as long as additional conditions outlined below are fulfilled.

In conjunction with this assumption one can be tempted to require that the particles be carried along by the flow.  Then there would be no hydrodynamic interactions whatsoever since particles would not exert any force on the fluid.  
In contrast, we assume that the particles are {\em held\/} in space (by optical tweezers, by inhomogeneities of the wall, by interactions with other particles in suspensions or meta-stable crystallites or by some other mechanism) while the fluid is flowing by and exerting a Stokes force on them.  The mechanism of such confinement and its limitations will be discussed in the discussion section.

Another condition heavily employed below also originates from the experimental procedure.  All measurements were conducted on colloidal suspensions enclosed between two microscope slides separated by a small gap (of the width of a few microns) so that the particles were confined in the vicinity of at least one of the slides.  Moreover, the strongest attraction was observed when particles were located sufficiently close to one of the slides.  Therefore, we consider particle interactions {\em near a plane wall\/}.  As it will be shown, both effects described below are absent in the infinite space.  

The last assumption used will be the absence of inertial effects or their negligible contribution, which is a reasonable approximation for the case of small Reynolds numbers (the numbers involved are of the order of $10^{-5}$--$10^{-6}$).  Gravity is also unimportant due to roughly equal densities of the material of the particles and the fluid, and so the only forces acting on the particles are of the hydrodynamic origin.  Thus, we consider {\em the hydrodynamic interactions of spatially fixed particles subject to slow flows (with small Reynolds numbers) in the semi-infinite space\/}.

\begin{figure}
\begin{center}
\includegraphics{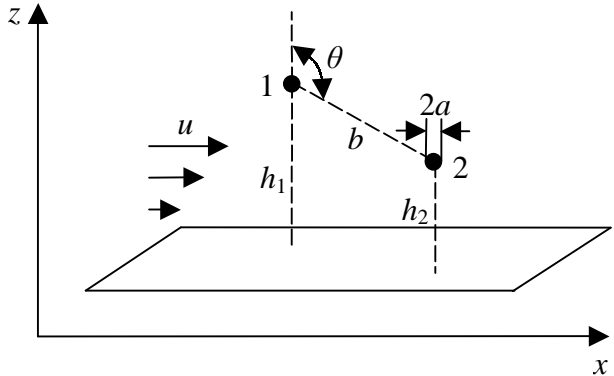}
\caption{Geometry of the problem.}
\label{fig1}
\end{center}
\end{figure}

In particular, consider particles 1 and 2 in the flow field above a plane wall.  Let particle 1 be at height $h_1$ above the plane and particle 2 be at $h_2$, and let the full (three-dimensional) distance between the particles be $b$ (Fig.~\ref{fig1}).  In agreement with the experimental conditions, we picture both particles as spheres; let their radius be $a$.  We do {\em not\/} assume any relation between $h$ and $b$, in particular they can be comparable or one of them can be much larger than the other.  However, we {\em do\/} assume that the size of the particles $a$ is much smaller than any other distance scale ($h$ or $b$), so that the particles can be thought of as point-like.  This last assumption is not strictly obeyed in the experiments (although particles {\em are\/} smaller than the separations between the objects), but it greatly simplifies the consideration and presumably does not change the qualitative character of the interaction potential.

We choose the origin of the system of coordinates at the location of particle 1.  Axis $z$ is directed away from the wall, so that the wall is defined by the equation $z = - h_1$.  Axis $x$ is directed along the velocity of the flow $\bf u$ (parallel to the wall).  Particle 2 has spherical coordinates $(b, \theta, \phi)$ or Cartesian coordinates $(b \sin\theta \cos\phi, b \sin\theta \sin\phi, b \cos\theta)$ with the standard choice of angles $\theta$ and $\phi$.  Then $h_2 = h_1 + b \cos\theta$.  (See Fig.~\ref{fig1} for geometry.)

\section{The shear flow effect}

The simplest effect in the system described above comes from the flow gradient near a plane wall (see {\em e.g.\ }Landau~\cite{landau}).  Sufficiently close to the wall the flow speed ${\bf u}$ must grow linearly with the distance to the wall:  ${\bf u} = \gamma h \hat x$, where $\gamma$ is the shear modulus.  Then the original flow field at the location of particle 1 creates a Stokes force on that particle ${\bf F}_1 = 6 \pi \eta a \gamma h_1 \hat x$, while at the location of particle 2 this force is ${\bf F}_2 = 6 \pi \eta a \gamma h_2 \hat x$, where $\eta$ is the fluid viscosity.  Obviously, the higher particle experiences a higher force from the fluid flowing by.  Now, if we define the apparent interaction force between the particles as the difference between the radial components of forces acting on each particle:  $F_r = {\bf F}_2 \cdot\hat r - {\bf F}_1 \cdot\hat r$ (so that the positive force corresponds to repulsion and the negative one corresponds to attraction), then the difference in forces exerted on the particles can be interpreted as either repulsion (if particle 2 is in the 1st or the 3rd quadrant of the $xz$-plane with respect to particle 1) or attraction (if particle 2 is in the 2nd or the 4th quadrant).  The exact result for the defined above interaction force is
\begin{equation}
F_r = 6 \pi \eta \gamma a b \cos\phi \cos\theta \sin\theta.
\label{force-trivial} \end{equation}
Of course, the interpretation of this force as repulsion or attraction does not mean that one of the particles acts on the other, but in the conditions of the experiment (where the potential is inferred from the measurements of Brownian motion for a fraction of a second) there is no simple way to distinguish whether particles interact directly or just move under the two external forces as if they interact.  

If the two particles are at the same height above the wall, then no apparent interaction force is present.  This can be seen both from expression~(\ref{force-trivial}) and from the fact that the Stokes forces exerted on both particles are exactly the same.  It can also be shown that there is no torque or force making particles leave the original height.

Expression~(\ref{force-trivial}) allows one to estimate the minimal value of the flow velocity necessary to produce the observed magnitude of the interaction force.  The maximum of the absolute value of angular dependence of the force is reached at $\theta = \pi / 4 \mbox{ or } 3 \pi / 4$ and $\phi = 0 \mbox{ or } \pi$; at this configuration $ | \cos\phi \cos\theta \sin\theta | = 1/2$.  Taking the typical values $F_r = 10^{-15}\mbox{ N}$, $a = 0.5 \times 10^{-6}\mbox{ m}$, $b = 2 \times 10^{-6}\mbox{ m}$, $\eta = 10^{-3}\mbox{ Pa}\cdot\mbox{s}$~\cite{crocker1, crocker2, grier4}, one can easily obtain:
\begin{equation}
\gamma \ge 0.1\mbox{ s}^{-1}.
\label{gamma-min} \end{equation}
Thus, flows as low as $0.1\mbox{ }\mu\mbox{m}/\mbox{s}$ (at height of $1\mbox{ }\mu\mbox{m}$ above the wall) can produce the observed magnitude of interactions.  These values of the velocity are below the experimentally detectable level, and such flows {\em can\/} exist in the system~\cite{grier4}.

Now, if there are two alternatives (attraction in 2nd and 4th quadrants and repulsion in 1st and 3rd quadrants), why would only attraction be observed?  A possible explanation will be deferred until the next section.

\section{Force propagation via viscous fluid}

Apart from the trivial effect described above, there is a more elaborate mechanism of particle interactions via viscous force propagating through the solvent.  If a point force $\bf F$ (``stokeslet'') is applied at point $\bf x$, then the velocity field perturbation $\bf v$ at point $\bf y$ is linearly related to the stokeslet magnitude:  $v_i({\bf y}) = H_{ij}({\bf y}-{\bf x}) F_j({\bf x})$, where $H_{ij}$ is known as the Oseen tensor (in the case of an extended source of force an integration over coordinates is required, so that the Oseen tensor is just a Green function for the velocity field).  The force exerted on a spherical particle at point $\bf y$ is then a Stokes force created by the velocity perturbation $\bf v$:  ${\bf f} = 6 \pi a \eta {\bf v}$.  In the case of two particles in viscous fluid the Stokes force ${\bf F}_1$ exerted by the original flow on particle 1 (given in the previous section, but taken with a minus sign, since now the particle, being held in place, acts on the fluid) will be the source of the perturbation of the velocity field ${\bf v}_2$ at the location of particle 2 and hence the source of the perturbation force ${\bf f}_2$ acting on particle 2, and vice versa.  Thus, the interaction force between the particles can again be defined as the difference between the radial components of forces acting on each particle:  $f_r = {\bf f}_2 \cdot\hat r - {\bf f}_1 \cdot\hat r$, but now this will be the true interaction force, the one propagating through the fluid.

\begin{figure}
\begin{center}
\includegraphics{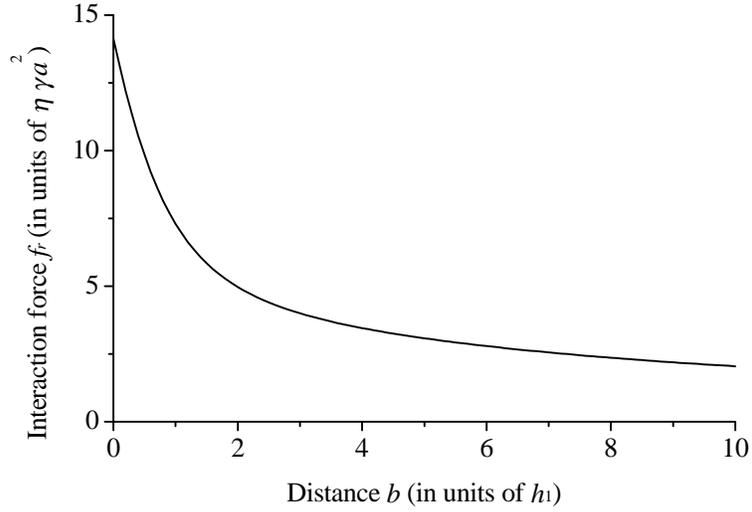}
\caption{Interaction force $f_r / (\eta \gamma a^2)$ as a function of distance $b/h_1$ for $\theta = \pi / 4$ and $\phi = 0$.}
\label{fig2}
\end{center}
\end{figure}

Expression for the Oseen tensor $H_{ij}$ in the infinite space is well known (see {\em e.g.\ }Happel and Brenner~\cite{happel}) and leads to the identical zero for the radial interaction force (the flow is uniform).  Thus, no hydrodynamic interactions are possible in the infinite space.  The Green function $H_{ij}$ for the semi-infinite space was constructed by Blake~\cite{blake}, although it can be alternatively derived by the method of Lorentz~\cite{lorentz}.  Calculation of the interaction force $f_r$ based on his result leads to the following expression:
\begin{equation}
f_r = 9 \pi \eta \gamma a^2 \cos\phi \cos\theta \sin\theta \left[ 1 - \frac{1 + 4 t_1 t_2 + 6 t_1 t_2 (t_1+t_2)^2}{(1 + 4 t_1 t_2)^{5/2}} \right],
\label{force-non-trivial} \end{equation}
where $t_1 \equiv h_1/b$ and $t_2 \equiv h_2/b = t_1 + \cos\theta$.  The asymptotics of the above result are:
\begin{equation}
f_r = 54 \pi \eta \gamma a^2 \frac{h_1}{b} \cos\phi \cos^2\theta \sin^3\theta \qquad\mbox{ if }h_1 \ll b
\label{hllb} \end{equation}
and
\begin{equation}
f_r = 9 \pi \eta \gamma a^2 \cos\phi \cos\theta \sin\theta \left( 1 - \frac{3b}{4h} \right)\qquad\mbox{ if }h \gg b,
\label{hggb} \end{equation}
where in the last line $h$ can be either $h_1$ or $h_2$.  Typical behavior of $f_r$ as a function of $1 / t_1 = b / h_1$ for $\theta = \pi / 4$ and $\phi = 0$ is shown on Fig.~\ref{fig2}.

Inspection of the result~(\ref{force-non-trivial}) indicates that the coefficient preceding this force is $a^2$, in contrast to the factor $a b$ in eq.~(\ref{force-trivial}).  However, far from the wall the angular dependences of the kinetic force~(\ref{force-trivial}) and the real force~(\ref{force-non-trivial}) are identical.  Thus, no new effect is present here:  although the force magnitude is different, the signage is the same --- attraction in 2nd and 4th quadrants and repulsion in 1st and 3rd quadrants.  So, expression~(\ref{force-non-trivial}) serves just as a correction to the main result~(\ref{force-trivial}) (since, by assumption, $a \ll b$).  It also gives the same order of magnitude for $\gamma$ as estimate~(\ref{gamma-min}) does.

\begin{figure}
\begin{center}
\includegraphics{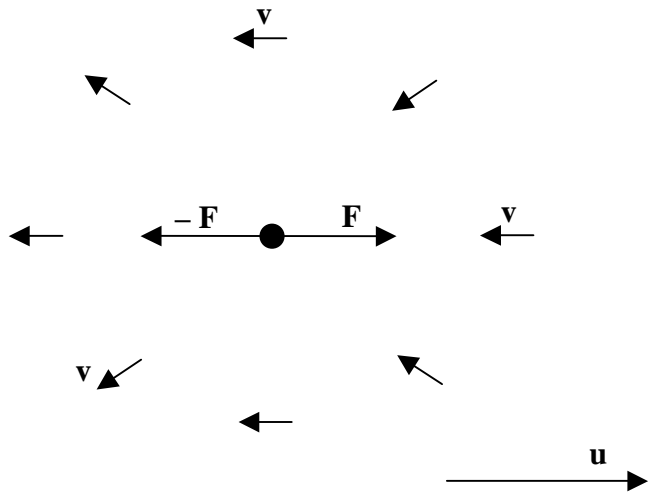}
\caption{Typical orientation of bulk flow velocity $\bf u$, Stokes force exerted by the fluid on the particle $\bf F$, Stokes force exerted by the particle on the fluid $-{\bf F}$, and velocity perturbation field ${\bf v}({\bf x})$ at some distance $b$ from the particle (infinite space).}
\label{fig3}
\end{center}
\end{figure}

Note also that $f_r$ of eq.~(\ref{force-non-trivial}) reinforces $F_r$ of eq.~(\ref{force-trivial}).  This is easy to understand from the fact that the perturbation velocity $\bf v$ (and hence the perturbation force ${\bf f}$) is in general oppositely directed with respect to the original Stokes force $\bf F$ acting from the fluid on the particle, or similarly directed with respect to the force $-{\bf F}$ from the particle on the fluid (see Fig.~\ref{fig3} for the infinite space), and is linearly proportional to the magnitude of the force ${\bf F}$.  Thus, for instance, if the radial projection ${\bf F}_1 \cdot\hat r$ is greater than ${\bf F}_2 \cdot\hat r$, then the negative radial projection $- {\bf f}_2 \cdot\hat r$ should in general be greater than $- {\bf f}_1 \cdot\hat r$, leading to the same signage of the above defined $F_r$ and $f_r$.

While the direct effect of the true hydrodynamic forces is relatively weak comparing to the kinetic effects of the previous section, they have an indirect effect that may explain the tendency towards observable attraction.  In particular, if the two particles are at the same height above the wall, then there is a force in vertical direction pushing each of the particles into the 2nd or the 4th quadrant with respect to the other particle:
\begin{equation}
f_z = - 54 \pi \eta \gamma a^2 \cos\phi \frac{t^4}{(1+4t^2)^{5/2}},
\label{fz} \end{equation}
where $t = h/b$ and $h = h_1 = h_2$.  For $h = b$, $\phi = 0$, $\gamma = \gamma_{min} = 0.1\mbox{  s}^{-1}$ and the same typical values of $a$ and $\eta$ as in estimate~(\ref{gamma-min}) the absolute value of this force is $7.6 \times 10^{-17}\mbox{ N}$, which is about 8\% of the typical value of $F_r$.  In general, for arbitrary $h_1$ and $h_2$, this effect can be described by the torque acting in the $\theta$-direction, which does not vanish at $\theta = \pi / 2$ (unlike in the effect of the previous section).  This torque is defined as $T = ({\bf f}_2 \cdot\hat\theta - {\bf f}_1 \cdot\hat\theta) b/2$, and a general expression for $T$ can be obtained on the same grounds as result~(\ref{force-non-trivial}):
\begin{equation}
T = \frac{9\pi}{4} \eta \gamma a^2 b \cos\phi \left[ \cos^2\theta - \frac{(1 + 4 t_1 t_2) \cos^2\theta + 6 t_1 t_2 (t_1+t_2)^2 \cos 2\theta}{(1 + 4 t_1 t_2)^{5/2}} \right].
\label{torque} \end{equation}
Thus, this torque indeed favors the attraction geometries by pushing particles into the configurations where attraction is created by the shear flow effect of the previous section.  Therefore, attraction indeed should be observed more often than repulsion if particles are positioned at the same height above the wall at the beginning of each experimental run.

\section{Discussion}

While the mechanism described in the previous section seems to lead to the correct results and be a feasible explanation to the observed attraction, there remains one question:  What holds the particles fixed in space?  As we mentioned earlier, the difference in fluid and particle velocities is the necessary condition for the existence of the Stokes forces producing all the above effects.  On the other hand, even if a particle is initially set at rest in the flow, while not being held by some external force, it acquires the velocity of the flow for relaxation time $\tau$ (defined by ${\bf v}_0(t) = {\bf u} \left( 1 - \exp(-t/\tau) \right)$) much shorter than the duration of each recording period ($1/30$~s), and hence it does not exert any force on the fluid for most of the recording time.  For instance, for a spherical particle of radius $a = 0.5 \times 10^{-6}\mbox{ m}$ and of density of the water ($\rho = 10^3\mbox{ kg/m}^3$) in a fluid of viscosity of the water ($\eta = 10^{-3}\mbox{ Pa}\cdot\mbox{s}$) this relaxation time is
\begin{equation}
\tau = \frac{2}{9} \frac{\rho a^2}{\eta} \approx 56\mbox{ ns}.
\label{tau} \end{equation}
Thus, the described effects {\em require\/} an external force that holds the particles at a non-zero velocity with respect to the flow.

The most probable candidate for this role could be a pair of optical tweezers used to trap the particles before each recording run.  One scenario is straightforward to consider.  While being held by the tweezers before each recording of their Brownian motion, particles are positioned to an attractive configuration by means of the torque described in the previous section.  After the holding potential of the tweezers is removed and the recording starts, the dominant shear flow effect influences the Brownian motion of the particles, and an attractive ``potential'' is registered.  Note that this does not even require any Stokes forces to be present --- just a drift in the flow with different velocities (so that the further from the wall particle moves faster) influences the Brownian motion in essentially the same manner.  Since the initial alignment was practically always chosen in the focal plane (so that both particles were at the same height above the wall) and {\em along\/} the longest dimension of the slides (so that one of the particles was always down the flow with respect to the other)~\cite{grier4}, one might speculate that the hydrodynamic effects proposed above had to produce apparent attraction in practically {\em all\/} the conducted experiments.  

However, in spite of general feasibility of such a mechanism, its estimated magnitude is discouraging.  In the particular conditions of the experiment the force gradient in the optical tweezers is of the order of $1\mbox{ pN/}\mu\mbox{m} = 10^{-6}\mbox{ N/m}$~\cite{grier4}, which leads to the forces of 3 or 4 orders of magnitude higher than force~(\ref{fz}).  Thus, particles are not easily displaceable by the hydrodynamic forces while being held by optical tweezers, and therefore the hydrodynamic mechanism appears too weak if the tweezers work as it is commonly understood.  (Note, however, that the estimate for the force~(\ref{fz}) was made for the {\em minimal\/} predicted value of the shear modulus;  higher flows lead to stronger hydrodynamic interaction.)  Therefore, optical tweezers, in spite of being a promising candidate, most likely cannot perform the role of the required external force.

Other candidates to this role might also be possible ({\em e.g.\ }forces due to the charges on the wall, due to inhomogeneities of the wall, or concerted mean-field-like forces at high-concentration systems like suspensions and meta-stable crystallites); however, we were not able to find a feasible scenario of how exactly these forces would keep the particles fixed (or at least at non-zero relative velocity with respect to the flow).  Thus, while the hydrodynamic mechanism certainly exists when such forces are present, the main difficulty in possible accounting for the attraction observed in the experiments like those of Grier {\em at al.} comes from identifying these forces.  At the same time, in the situations where there are such forces, the hydrodynamic mechanism considered here is always at work (and this should be kept in mind while conducting high-sensitivity force measurements in colloidal systems).  One example where the required forces exist can be the systems of meta-stable crystallites~\cite{larsen1, larsen2}.  A self-sustaining mechanism can be employed here:  The attractive force holds particles together, creating thereby the necessary condition for the described effect (flow past {\em fixed\/} particles --- assuming some flow is present), which in its turn leads to the existence of attraction.  Further work would be required to account in detail for such a self-sustaining mechanism.  Here we only notice that while the concerted forces acting on a particular particle can be mean-field-like, the whole system of particles cannot be held at rest by only the interaction forces between the particles in that system, as in this case the whole system would be carried along by the flow.  Therefore, there should be some external forces different from interactions with the neighboring particles and acting at least on the boundary layer of the crystallite system.
 
As it is also apparent from our consideration, flow is not the only possible source of the described effect.  What is necessary is the force exerted by the particles onto the fluid.  This can be achieved not only by making a fluid flow by the fixed particles but also by making particles move in the stationary fluid.  Thus, for instance, particles can be dragged by some external force in such a way that the further from the wall particle experiences higher force; this should lead to effectively the same result for the interaction potential.  Essentially, this has a lot in common with the mechanism of Squires and Brenner~\cite{squires}, where particles are dragged away from the wall by the electromagnetic repulsive force.

Note that presence of a second wall, positioned symmetrically with respect to the interacting particles, leads to canceling of all considered effects (both force and torque).  Therefore, another restriction on the possible experimental geometries where this hydrodynamic mechanism could be responsible for interactions is that the particles must be located away from the center plane of the suspension sample confined by the microscope slides.  In practice, however, this restriction is easily avoided due to a slight difference in particle and fluid densities, so that the particles are pulled closer to the lower wall by the gravity (if their density is higher) or pushed closer to the upper wall by the Archimedes force (if their density is lower).

The proposed mechanism immediately suggests several possible experimental tests on the presented effects of flow.  Most of these tests originate from the following list of the predicted properties:

\begin{enumerate}
\item The simplest hydrodynamic effect originates from the velocity gradient near a surface and can lead to the apparent attraction or repulsion (although no direct interaction exists).  Both outcomes are equally probable with randomly chosen initial geometries.
\item The true hydrodynamic interactions propagating through the viscous fluid {\em do\/} exist, but they lead to the same qualitative dependence on the geometry --- attraction in 2nd and 4th quadrants and repulsion in 1st and 3rd quadrants.  Nevertheless, there exists a mechanism favoring attraction geometries.
\item The flows necessary to produce any hydrodynamic effects can be as low as $0.1\mbox{ }\mu\mbox{m}/\mbox{s}$ (more generally, $\gamma \ge 0.1\mbox{ s}^{-1}$).
\end{enumerate}

Different initial geometries fixed by optical tweezers can be tried to check if the interaction strength changes depending on the position of the particles.  This requires only relocation of principal potential minima of the tweezers, which can be relatively easily achieved by refocusing or rotating them.  If both attractive and repulsive results are recorded (or if a stronger interaction produced by some other mechanism gets modified accordingly), then the effect is at work.

One can also look explicitly for the flow.  This flow should be easily detectable if particles are allowed to drift freely for a sufficient interval of time.  The flow could also be deliberately induced at higher levels to increase the magnitude of the hydrodynamic force and check that the mechanism works as predicted.

\vspace{3ex}

{\small  The mechanism discussed here and the method to estimate its magnitude were proposed by T.A.~Witten, who also provided ongoing advice during this work.  D.G.~Grier as well as other people of his group (especially J.~Plewa and S.~Behrens) supplied essential experimental input and a number of ideas about the proposed mechanism and its feasibility.  This work was supported in part by the MRSEC Program of the National Science Foundation under Award Number DMR-9808595.}

\end{document}